\documentclass[12pt]{article}
\usepackage{amssymb}
\usepackage{amsmath}
\usepackage{diagbox}
\usepackage[ruled,vlined]{algorithm2e}

\numberwithin{equation}{section}
\def \bes{\begin{eqnarray}}
\def \ees{\end{eqnarray}}
\def \bns{\begin{eqnarray*}}
\def \ens{\end{eqnarray*}}

\newcommand{\be}{\begin{eqnarray}}
\newcommand{\ee}{\end{eqnarray}}
\newcommand{\non}{\nonumber}

\newcommand{\tr}{\mathop{\rm tr}\nolimits}

\begin{document}

\begin{titlepage}
\strut\hfill UMTG--278
\vspace{.5in}
\begin{center}

\LARGE Singular solutions, repeated roots and completeness\\
for higher-spin chains\\
\vspace{1in}
\large Wenrui Hao \footnote{
Mathematical Biosciences Institute, The Ohio State University, 1735
Neil Avenue, Columbus, OH 43210 USA, hao.50@osu.edu}, 
Rafael I. Nepomechie \footnote{
Physics Department,
P.O. Box 248046, University of Miami, Coral Gables, FL 33124 USA,
nepomechie@physics.miami.edu}
and Andrew J. Sommese \footnote{
Department of Applied and Computational Mathematics and Statistics,
University of Notre Dame, 153 Hurley Hall, IN 46556 USA,
sommese@nd.edu}
\\[0.8in]
\end{center}

\vspace{.5in}

\begin{abstract}
We investigate the completeness of the solutions of the Bethe
equations for the integrable spin-$s$ isotropic (XXX) spin chain with
periodic boundary conditions.  Solutions containing the exact string
$i s, i (s-1), \ldots, -i(s-1), -is$ are singular.  For $s>1/2$, there
exist also ``strange'' solutions with repeated roots, which
nevertheless are physical (i.e., correspond to eigenstates of the
Hamiltonian).  We derive conditions for the singular solutions and the
solutions with repeated roots to be physical.  We formulate a
conjecture for the number of solutions with pairwise distinct roots in
terms of the numbers of singular and strange solutions.  Using
homotopy continuation, we solve the Bethe equations numerically for
$s=1$ and $s=3/2$ up to 8 sites, and find some support for the
conjecture.  We also exhibit several examples of strange solutions.
\end{abstract}

\end{titlepage}

\setcounter{footnote}{0}

\section{Introduction}\label{sec:intro}

The spin-1/2 isotropic Heisenberg quantum spin chain, which is well
known to be integrable and therefore solvable by Bethe ansatz (see
e.g. \cite{Faddeev:1996iy}), has an integrable spin-$s$ generalization
($s=1/2, 1, 3/2, \ldots$) \cite{Zamolodchikov:1980ku, Kulish:1981gi, 
Kulish:1981bi, A.:1982zz,
Babujian:1983ae}.  Since there is a spin-$s$ spin at each of the $N$ 
sites, the Hilbert space is $({\cal C}^{2s+1})^{\otimes N}$.
The Hamiltonian is given (up to an additive 
constant) by \cite{Faddeev:1996iy, Babujian:1983ae}
\be
H_{s} = \sum_{n=1}^{N}Q_{2s}(\vec s_{n} \cdot \vec s_{n+1}) \,, 
\qquad \vec s_{N+1} \equiv \vec s_{1} \,, 
\label{Hamiltonian}
\ee
where $\vec s$ are the spin-$s$ generators of $su(2)$, $\vec s_{n}$ 
denotes the spin operators at site $n$, $Q_{2s}(x)$ is
a polynomial of degree $2s$ given by
\be
Q_{2s}(x) = \sum_{j=1}^{2s} h(j) \prod_{\scriptstyle{l \ne j}\atop 
\scriptstyle{l=0}}^{2s} \frac{x-x_{l}}{x_{j}-x_{l}} \,, \qquad 
x_{l} = \frac{1}{2}l(l+1)-s(s+1) \,,
\ee
and $h(j)$ is the $j^{th}$ harmonic number,
\be
h(j)=\sum_{k=1}^{j}\frac{1}{k} \,.
\ee 
For example, for $s=1$, $Q_{1}(x)=\frac{1}{4}(x-x^{2})+\frac{3}{2}$.
The eigenvalues of the Hamiltonian (\ref{Hamiltonian}) are given by \cite{Faddeev:1996iy,
A.:1982zz, Babujian:1983ae}
\be
E_{s} = -\sum_{k=1}^{M}\frac{s}{\lambda_{k}^{2}+s^{2}} \,,
\label{energy}
\ee
where $\{\lambda_{1} \,, \ldots \,, \lambda_{M}\}$ satisfy the 
spin-$s$ Bethe equations
\be
\left(\lambda_{k} + i s\right)^{N}
\prod_{\scriptstyle{j \ne k}\atop \scriptstyle{j=1}}^M
(\lambda_{k} - \lambda_{j} - i)
= \left(\lambda_{k} - i s \right)^{N}
\prod_{\scriptstyle{j \ne k}\atop \scriptstyle{j=1}}^M
(\lambda_{k} - \lambda_{j} + i) \,, \non \\
\qquad k = 1 \,, 2\,, \ldots \,, M \,, \qquad
M = 0\,, 1\,, \ldots\,,  s N \,.
\label{BAE}
\ee
The corresponding Bethe states are given by
\be
|\lambda_{1} \,,
\ldots \,, \lambda_{M} \rangle = \prod_{j=1}^{M} B(\lambda_{j}) 
|0\rangle \,, \qquad |0\rangle= {\scriptscriptstyle\left(\begin{array}{c}1\\ 0 \\ 
\vdots\\0 \end{array}\right)}^{\otimes N} \,,
\label{vec}
\ee
where the creation operator $B(\lambda)$ is defined in Appendix
\ref{sec:singular}, and $|0\rangle$ is the reference state with all
spins ``up'' (eigenstates of $s^{z}$ with eigenvalue $s$).  These
states are eigenvectors of the Hamiltonian (\ref{Hamiltonian}) with eigenvalue
(\ref{energy}) as well as $su(2)$ highest-weight states
\be
S^{+} |\lambda_{1} \,,
\ldots \,, \lambda_{M} \rangle = 0 \,, \qquad S^{\pm} = S^{x} \pm i 
S^{y}\,, \qquad \vec S = \sum_{n=1}^{N} \vec s_{n} \,,
\ee
and are therefore eigenstates of $\vec S^{2}$ and $S^{z}$ with
corresponding eigenvalues $S(S+1)$ and $m$ (respectively) with
\be
S=m=s N - M \,.
\label{quantumno}
\ee
The states with lower values of $m$ can be
obtained by acting on the Bethe states (\ref{vec}) with the spin
lowering operator $S^{-}$.

For $s>1/2$, the models (\ref{Hamiltonian}) are perhaps not realizable
experimentally.  Nevertheless, these models have interesting
connections to the level $k=2s$ Wess-Zumino-Witten model and RSOS
models (see e.g. \cite{Faddeev:1996iy, Affleck:1988px,
Reshetikhin:1990jn} and references therein).

A still-unanswered question is whether the Bethe ansatz solution is
``complete''; i.e., whether the Bethe equations (\ref{BAE}) have the correct
number of solutions to account for all $(2s+1)^{N}$ eigenstates of the
Hamiltonian.

For the $s=1/2$ chain, this problem was already discussed by Bethe in
his original paper \cite{Bethe:1931hc}, and was subsequently further
investigated by others (see e.g. \cite{Faddeev:1996iy, Takahashi1971,
Kirillov:1985, Langlands:1995, Langlands:1997, Fabricius:2000yx,
Baxter:2001sx, Mukhin:2009}).  Part of the difficulty of this problem
is due to the existence of so-called singular (or exceptional)
solutions of the Bethe equations, i.e. solutions that contain $\pm
i/2$, and which do not necessarily correspond to eigenstates of the
Hamiltonian (see e.g. \cite{Avdeev:1985cx, Essler:1991wd,
Siddharthan:1998, Noh:2000, Beisert:2003xu, Beisert:2004hm,
Hagemans:2007, Arutyunov:2012tx, Nepomechie:2013mua}).  Another
difficulty is that the possibility of repeated Bethe roots has not
been ruled out.

We recently formulated, for this $s=1/2$ case, a precise conjecture
\cite{Hao:2013jqa} for the number of solutions of the Bethe equations
with pairwise distinct roots, in terms of numbers of singular
solutions.  Using homotopy continuation methods (see e.g. \cite{SW2, BHSW1}
and references therein), we solved the Bethe equations numerically for
values of $N$ up to 14, and found perfect agreement with this
conjecture.  Its meaning is that the Bethe equations generally have
too many solutions with pairwise distinct roots; but after discarding
the unphysical singular solutions, there remain exactly the right
number of solutions to account for all $2^{N}$ eigenstates of the
Hamiltonian.

Our goal in the present work is to extend our $s=1/2$ results
\cite{Hao:2013jqa} to arbitrary values of $s$. We find that the 
notion of singular solution neatly generalizes to the spin-$s$ case: 
a solution is singular if it contains an exact string of length $2s+1$ 
centered at the origin, i.e.
\be
\{ i s\,, i(s-1)\,, \ldots \,, -i(s-1)\,, -i s \} \,.
\label{singular}
\ee
However, in contrast with the $s=1/2$ case, we find that for $s>1/2$
there also exist ``strange'' solutions: solutions of the Bethe
equations that have repeated roots, which nevertheless produce
eigenstates of the Hamiltonian.  Consequently, checking
completeness is significantly more complicated.

The outline of this paper is as follows.  In Section
\ref{sec:completeness} we formulate a conjecture for the number of
solutions of the Bethe equations (\ref{BAE}) with pairwise distinct
roots in terms of the numbers of singular and strange solutions.  In
Section \ref{sec:numerical} we present some numerical support for this
conjecture for the cases $s=1$ and $s=3/2$.  We also exhibit several
examples of strange solutions.  Our analysis relies on important
conditions for singular and strange solutions to be physical, the
derivations of which are presented in Appendices \ref{sec:singular}
and \ref{sec:strange}, respectively.  A brief discussion of our
results is given in Section \ref{sec:discuss}.

\section{The completeness conjecture}\label{sec:completeness}

As in the $s=1/2$ case \cite{Hao:2013jqa}, we denote by ${\cal
N}(N,M)$ the number of solutions of the Bethe equations (\ref{BAE})
for given values of $N$ and $M$ with pairwise distinct roots
(i.e., $\lambda_{j} \ne \lambda_{k}$ for $j\ne k$).  We always count
solutions up to permutations of the roots: if $\{\lambda_{1} \,,
\ldots \,, \lambda_{M}\}$ is a solution, then any permutation of these
$\lambda_{k}$'s is not counted as a separate solution.

A naive conjecture for ${\cal N}(N,M)$ can be obtained on the basis of
the Clebsch-Gordan theorem. Indeed, the $N$-fold tensor product of 
spin-$s$ representations decomposes into a direct sum of 
(irreducible) spin-$S$
representations,
\be
\underbrace{{\bf s}\otimes \cdots \otimes {\bf s}}_{N} =
\bigoplus_{S=S_{\rm min}}^{s N} n(N,S)\, {\bf S}\,,
\label{CG}
\ee
where $S_{\rm min}=0$ if $s N$ is an integer, and $S_{\rm min}=\frac{1}{2}$ if $s 
N$ is a half-odd integer. The multiplicity of the spin-$S$ representation, 
which we denote by $n(N,S)$, is computed in the Appendix of
\cite{Mendonca:2013} \footnote{This result is derived using an 
expression for $C_{m}(n,k)$ defined by \cite{Bollinger:1993}
\be
(1+x+x^{2}+\cdots +x^{m-1})^{n} = \sum_{k=0}^{(m-1)n}C_{m}(n,k) 
x^{k}\,. \non
\ee
However, there is a misprint in
Eq. (3) of \cite{Bollinger:1993}; the correct result is
\be
C_{m}(n,k) =\sum_{j=0}^{n}{}^{'}(-1)^{j}{n\choose j}{n-1+k-m j \choose k-m 
j} \,, \non
\ee
where the sum is restricted so that $k - m j \ge 0$. An alternative 
expression for the multiplicities is given by Lemma 6 in
\cite{Kirillov:1985}.}
\be
n(N,S) = b(N,S) - b(N,S+1) \,, 
\label{multiplicities}
\ee
where
\be
b(N,r) = \sum_{k=0}^{N}{}^{'}(-1)^{k} {N \choose k} {(s+1)N+r-(2s+1)k -1\choose 
sN+r -(2s+1)k} \,,
\label{multiplicities2}
\ee 
and the sum is restricted so that $sN+r - (2s+1)k \ge 0$. In view of 
(\ref{CG}), it is necessary that
\be
\sum_{S=S_{\rm min}}^{s N} (2S+1)\, n(N,S) = (2s+1)^{N} \,.
\ee 
Since $M=sN- S$ (\ref{quantumno}) and the values of $S$ range from
$S_{\rm min}$ to $sN$, it follows that the values of $M$ range from 0
to $[sN]$, as anticipated in (\ref{BAE}).  If we naively assume that
every solution of the Bethe equations with pairwise distinct roots
produces, via (\ref{vec}), an $su(2)$ highest-weight eigenstate of the
Hamiltonian, then the number of solutions of the Bethe equations with
$M$ pairwise distinct roots should be given by the multiplicity of 
representations with spin $S=sN-M$.  That is,
\be
{\cal N}(N,M) \stackrel{\rm ?}{=} n(N,sN-M) \,.
\label{naive}
\ee
A proof of (\ref{naive}) (based on the string hypothesis
\cite{Faddeev:1996iy, Bethe:1931hc, Takahashi1971}, which requires $N
\rightarrow \infty$) was proposed in \cite{Kirillov:1985}.  Related
arguments were given in \cite{A.:1982zz, Babujian:1983ae}.

This naive conjecture is wrong for two reasons.  First, as discussed
in detail in Appendix \ref{sec:singular}, there exist unphysical
singular solutions with pairwise distinct roots, which do {\em not} produce
eigenstates of the Hamiltonian.  Second, as discussed
in Appendix \ref{sec:strange}, there exist ``strange''
solutions with repeated roots, which {\em do} produce eigenstates of the
Hamiltonian. Correcting for these two effects, we arrive at the 
following conjecture:
\be
{\cal N}(N,M) -{\cal N}_{s}(N,M) + {\cal N}_{sp}(N,M) + {\cal N}_{strange}(N,M) 
=  n(N,sN-M) \,, 
\label{conjecture}
\ee
where ${\cal N}_{s}(N,M)$ is the number of solutions of the Bethe
equations (\ref{BAE}) with pairwise distinct roots that are singular 
(i.e., that contain (\ref{singular})); and ${\cal N}_{sp}(N,M)$
is the number of such singular solutions that satisfy (\ref{physical})
\be
\left[(-1)^{2s} \prod_{k=2s+2}^{M}\left( \frac{\lambda_{k} + 
is}{\lambda_{k} - is}\right) \right]^{N} = 1 \,,
\label{physical2}
\ee
and hence are physical.  (As explained in Appendix 
\ref{sec:singular}, singular solutions require regularization; and the
requirement that the corresponding Bethe states be eigenstates of the 
Hamiltonian leads to generalized Bethe equations, which in turn lead
to the constraint (\ref{physical2}). This constraint reduces for $s=1/2$ to the one 
recently found in \cite{Nepomechie:2013mua}.)
Note that ${\cal N}_{s} - {\cal 
N}_{sp}$ is the number of unphysical singular solutions.
Finally, ${\cal N}_{strange}(N,M)$ is the number of
strange solutions, whose roots are {\em not}
pairwise distinct but which nevertheless produce
eigenstates of the Hamiltonian. 

In principle, there are many possible types of solutions with
repeated roots; and each type should have its own physicality
condition.  Therefore, in practice, it is difficult to know if one has
found all the strange solutions. This, in turn, makes the conjecture 
(\ref{conjecture}) difficult to check. However, since ${\cal 
N}_{strange}(N,M) \ge 0$, this conjecture 
implies the inequality
\be
{\cal N}(N,M) -{\cal N}_{s}(N,M) + {\cal N}_{sp}(N,M)  
\le  n(N,sN-M) \,, 
\label{inequality}
\ee
which is much easier to check.  Its physical meaning is that the Bethe
equations (\ref{BAE}) generally have too many solutions with pairwise
distinct roots; but after discarding the unphysical singular
solutions, there may not remain sufficiently many solutions with
pairwise distinct roots to account for all $(2s+1)^{N}$ eigenstates of
the Hamiltonian (\ref{Hamiltonian}).  The equality (\ref{conjecture})
further asserts that any such deficit is made up by certain solutions
with repeated roots.

\section{Numerical results}\label{sec:numerical}

Using homotopy continuation with Bertini \cite{BHSW1} (see
\cite{Hao:2013jqa} and references therein for further details), we
have solved the Bethe equations (\ref{BAE}) numerically for $s=1$ and
$s=3/2$ with $N=2, 3, \ldots, 8$.  The results are presented in a set
of supplemental tables \cite{tables}, and are summarized below.

\subsection{The case $s=1$}

The results for $s=1$ are summarized in Table \ref{table:spin1}.  For each set of values
$(N,M)$, we report a set of five integers:
\be
({\cal N}\,, {\cal N}_{s}\,, {\cal N}_{sp}\,, {\cal N}_{strange}\,;
{\cal N}-{\cal N}_{s}+{\cal N}_{sp}+{\cal N}_{strange}) \,, \non
\ee
where ${\cal N}$ is the number of solutions of the Bethe equations
with pairwise distinct roots; ${\cal N}_{s}$ is the number of singular
solutions (i.e., that contain $\pm i, 0$) with pairwise distinct roots; ${\cal N}_{sp}$ is the number
of such singular solutions that are physical (i.e., that satisfy 
(\ref{physical2}) with $s=1$); and ${\cal N}_{strange}$
is the number of solutions with repeated roots that are physical.
The quantities ${\cal N}-{\cal N}_{s}+{\cal N}_{sp}+{\cal N}_{strange}$ in
all the entries of Table \ref{table:spin1} coincide with $n(N,N-M)$, 
in perfect agreement with the conjecture (\ref{conjecture}).

\begin{table}[htb]
    \tiny
  \centering
  \begin{tabular}{|l|c|c|c|c|c|c|c|c|}\hline
    \diagbox[width=2.5em]{$N$}{$M$} & 1 & 2 & 3 & 4 & 5 & 6 & 7 & 8 \\
    \hline
     2 & (1,0,0,0; 1) & (1,0,0,0; 1) & & & & & &\\
     3 & (2,0,0,0; 2) & (3,0,0,0; 3) & (1,1,1,0; 1) & & & & &\\
     4 & (3,0,0,0; 3) & (6,0,0,0; 6) & (6,1,1,0; 6) & (4,2,0,1; 3)& & 
     & &\\
     5 & (4,0,0,0; 4) & (10,0,0,0; 10) & (15,1,1,0; 15) & (19,4,0,0; 
     15) & (14,10,2,0; 6) & & &\\
     6 & (5,0,0,0; 5) & (15,0,0,0; 15) & (29,1,1,0; 29) & (43,4,0,1; 
     40) & (48,15,3,0; 36) & (41,28,0,2; 15) & &\\
     7 & (6,0,0,0; 6) & (21,0,0,0; 21) & (49,1,1,0; 49) & (90,6,0,0; 
     84)& (123,21,3,0; 105) & (141,50,0,0; 91)& (120,90,6,0; 36) &\\
     8 & (7,0,0,0; 7) & (28,0,0,0; 28) & (76,1,1,0; 76) & (159,6,0,1; 
     154) & (262,28,4,0; 238) & (351,74,0,3; 280) & (384,161,9,0; 
     232) & (345,260,0,6; 91) \\
    \hline
   \end{tabular}
   \caption{\small The values $({\cal N}\,, {\cal N}_{s}\,, {\cal 
   N}_{sp}\,, {\cal N}_{strange}\,;
{\cal N}-{\cal N}_{s}+{\cal N}_{sp} + {\cal N}_{strange})$ for $s=1$ 
and given values
of $N$ and $M$.}
   \label{table:spin1}
\end{table}

We observe that, up to $N=8$, all the strange solutions contain 
$\pm i\,, 0\,, 0$; and the remaining $M-4$ roots satisfy (\ref{strangespin1})
\be
\prod_{j=5}^{M}\left(\frac{\lambda_{j}+2i}{\lambda_{j}-2i}\right) = 
(-1)^{N} \,,
\ee
which is a condition for singular solutions of this particular type to be 
physical.
These solutions are listed in Table 
\ref{table:strangespin1}.\footnote{In \cite{Avdeev:1985cx},
the solution for $N=M=4$ was found (see Eqs. (40), (41)), and the 
corresponding Bethe vector was  
asserted to be the first and (at that time) only example with equal 
roots. The two solutions for $N=M=6$ were also noted in \cite{Avdeev:1985cx}
(see Eq. (20)), but were derived using the erroneous Eq. 
(19). See also footnote \ref{footnote:AV}.} It 
remains an open question whether there exist additional types of strange solutions 
for $s=1$ and $N>8$.

\begin{table}[htb]
  \centering
  \begin{tabular}{|c|c|c|}\hline
      N & M & $\lambda_{1}, \ldots, \lambda_{M}$ \\
    \hline
    4 & 4 & $\pm i$\,, 0\,, 0 \\
    \hline
    6 & 4 & $\pm i$\,, 0\,, 0 \\
    \hline
    6 & 6 & $\pm i$\,, 0\,,  0\,, $\pm 0.474498 i$\\
    6 & 6 & $\pm i$\,, 0\,, 0\,,  $\pm 2.10749 i$\\
    \hline
    8 & 4 & $\pm i$\,, 0\,, 0\\
    \hline
    8 & 6 & $\pm i$\,, 0\,, 0\,, $\pm 0.497637 i$\\
    8 & 6 & $\pm i$\,, 0\,, 0\,, $\pm 2.0095 i$ \\
    8 & 6 & $\pm i$\,, 0\,,  0\,, $\pm 1$\\
    \hline
    8 & 8 & $\pm i$\,, 0\,,  0\,, $\pm 0.833228$\,,  
    $\pm 2.06462 i$\\
    8 & 8 & $\pm i$\,, 0\,,  0\,, $\pm 2.00281 i$\,, 
     $\pm 3.53743 i$\\
    8 & 8 & $\pm i$\,, 0\,,  0\,, $\pm 0.500018 i$\,, 
    $\pm 1.49494 i$\\
    8 & 8 & $\pm i$\,, 0\,, 0\,, $0.15263  \pm  0.474611 i$\,,   
    $-0.15263 \pm 0.474611 i$\\
    8 & 8 & $\pm i$\,, 0\,,  0\,, $0.594869 \pm 0.532284 i$\,, $-0.594869 \pm 0.532284 i$\\
    8 & 8 & $\pm i$\,, 0\,,  0\,, $\pm 0.498746 i$\,, $\pm 2.13944 i$\\
     \hline
   \end{tabular}
   \caption{Strange solutions for $s=1$ up to $N=8$.}
 \label{table:strangespin1}
\end{table}

\subsection{The case $s=3/2$}

The results for $s=3/2$ are summarized in Tables \ref{table:spin32a}
and \ref{table:spin32b}.  Here the singular solutions contain $\pm
3i/2\,, \pm i/2$; and the physical singular solutions satisfy
(\ref{physical2}) with $s=3/2$.  Up to $N=7$, the quantities ${\cal N}-{\cal
N}_{s}+{\cal N}_{sp}+{\cal N}_{strange}$ in all the entries of these
tables coincide with $n(N,\frac{3N}{2}-M)$, in agreement
with the conjecture (\ref{conjecture}).

Our results for $N=8$ and $M=6, 8, 10, 12$ are inconclusive.  In order
to have agreement with the conjecture (\ref{conjecture}), a small
number of strange solutions (as indicated by the entries marked with
``?''  in Tables \ref{table:spin32a} and \ref{table:spin32b}) should
exist for those cases.  Unfortunately, we have not managed to identify
those solutions.  We speculate that those missing solutions are
strange singular solutions of a new type, containing both an exact 4-string
($\pm 3i/2\,, \pm i/2$) {\it and} an exact 2-string ($\pm i/2$).
However, as discussed in Section \ref{subsec:strange2}, we have not
been able to derive the consistency condition(s) for such solutions
(i.e., the conditions for the corresponding Bethe states to be
eigenstates of the Hamiltonian).  Hence, we can neither confirm nor 
contradict the conjecture for $N=8$ and $M=6, 8, 10, 12$.  The
inequality (\ref{inequality}) is nevertheless satisfied for these
cases.

\begin{table}[htb]
    \tiny
  \centering
  \begin{tabular}{|l|c|c|c|c|c|c|}\hline
    \diagbox[width=2.5em]{$N$}{$M$} & 1 & 2 & 3 & 4 & 5 & 6  \\
    \hline
     2 & (1,0,0,0; 1) & (1,0,0,0; 1) & (1,0,0,0; 1) & & & \\
     3 & (2,0,0,0; 2) & (2,0,0,1; 3) & (4,0,0,0; 4) & (3,1,0,0; 2) & & \\
     4 & (3,0,0,0; 3) & (6,0,0,0; 6) & (10,0,0,0; 10) & (11,1,1,0; 
     11) & (11,3,1,0; 9) & (8,6,2,0;4)\\
     5 & (4,0,0,0; 4) & (10,0,0,0; 10) & (20,0,0,0; 20) & (31,1,0,0; 
     30) & (40,4,0,0; 36) & (44,10,0,0; 34) \\
     6 & (5,0,0,0; 5) & (13,0,0,2; 15) & (35,0,0,0; 35) & (64,1,1,0; 
     64) & (100,5,1,0; 96) & (132,15,3,0; 120) \\
     7 & (6,0,0,0; 6) & (21,0,0,0; 21) & (56,0,0,0; 56) & (120,1,0,0; 
     119)& (216,6,0,0; 210) & (336,21,0,0; 315)\\
     8 & (7,0,0,0; 7) & (28,0,0,0; 28) & (84,0,0,0; 
     84) & (202,1,1,0; 202) & (412,7,1,0; 406) & (723,28,4,0 [1?]; 
     699) \\
     \hline
   \end{tabular}
   \caption{\small The values $({\cal N}\,, {\cal N}_{s}\,, {\cal 
   N}_{sp}\,, {\cal N}_{strange}\,;
{\cal N}-{\cal N}_{s}+{\cal N}_{sp} + {\cal N}_{strange})$ for $s=3/2$ 
and given values
of $N$ and $M$ (part 1).}
   \label{table:spin32a}
\end{table}
\begin{table}[htb]
    \tiny
  \centering
  \begin{tabular}{|l|c|c|c|c|c|c|}\hline
    \diagbox[width=2.5em]{$N$}{$M$} & 7 & 8 & 9 & 10 & 11 & 12  \\
    \hline
     2 &   &  &  & & & \\
     3 &  &  &  &  & & \\
     4 &  &  &  &  &  & \\
     5 & (40,20,0,0; 20) &  &  &  &  &  \\
     6 & (152,35,3,0; 120) & (150,65,5,0; 90) & (130,101,5,0; 34) &  &  &  \\
     7 & (456,56,0,0; 400) & (545,119,0,0; 426) & (580,216,0,0; 364) 
     & (543,333,0,0; 
     210)&  & \\
     8 & (1124,84,4,0; 1044) & (1544,203,9,0 [1?]; 1350) & (1909,413,9,0; 
     1505) & (2107,728,16,0 [5?]; 1395) & (2112,1128,16,0; 1000) & 
     (1893,1554,22,0 [3?]; 
     361) \\
     \hline
   \end{tabular}
   \caption{\small The values $({\cal N}\,, {\cal N}_{s}\,, {\cal 
   N}_{sp}\,, {\cal N}_{strange}\,;
{\cal N}-{\cal N}_{s}+{\cal N}_{sp} + {\cal N}_{strange})$ for $s=3/2$ 
and given values
of $N$ and $M$ (part 2).}
   \label{table:spin32b}
\end{table}

We have found strange solutions for $N=3, 6$ with $M=2$, which are 
listed in Table \ref{table:strangespi32}. These solutions are not 
singular, and therefore obey the consistency conditions for strange 
non-singular solutions discussed in Sec. \ref{sec:strangenonsing}, 
namely (\ref{extrastrangenonsingular}) with $k=2$. For the $M=2$ case 
considered here, these equations reduce to 
\be
\left(\frac{\mu-i s}{\mu+is} \right)^{N} = -1 \,, \qquad 
\frac{\partial}{\partial \mu} \left(\frac{\mu-i s}{\mu+is} 
\right)^{N} = -4i \,.
\ee 
The 
solution $\{0\,, 0\}$ for $s=3/2\,, N=3$ is particularly noteworthy, 
as it is surely the simplest 
physical solution of the Bethe equations with 
repeated roots.

\begin{table}[htb]
  \centering
  \begin{tabular}{|c|c|c|}\hline
      N & M & $\lambda_{1}, \ldots, \lambda_{M}$ \\
    \hline
    3 & 2 & 0\,, 0 \\
    \hline
    6 & 2 & 1.5\,, 1.5 \\
    6 & 2 & -1.5\,, -1.5  \\
    \hline
       \end{tabular}
   \caption{Strange solutions for $s=3/2$ up to $N=7$.}
 \label{table:strangespi32}
\end{table}

\section{Discussion}\label{sec:discuss}

Previous studies \cite{A.:1982zz, Babujian:1983ae, Kirillov:1985} of
the completeness of the solutions of the spin-$s$ Bethe equations
(\ref{BAE}) have relied on the string hypothesis \cite{Faddeev:1996iy,
Bethe:1931hc, Takahashi1971}, which requires $N \rightarrow \infty$
and whose status remains unclear (see below).

We have initiated here instead an ab initio investigation of the
completeness problem.  To this end, we have formulated a conjecture
(\ref{conjecture}) for the number of solutions with pairwise distinct
roots in terms of the numbers of singular and strange solutions.  We
have tested this conjecture for small values of $s$ and $N$.  For
$s=1$, there is perfect agreement with the conjecture up to $N=8$
(i.e., as far as we have checked).  For $s=3/2$, there is perfect
agreement with the conjecture up to $N=7$.  However, for $N=8$ and
$M=6, 8, 10, 12$, our results are inconclusive.  We expect that the
conjecture will continue to hold for these cases, and that the
observed discrepancies will be resolved by the existence of a new type
of strange singular solution, which remains to be confirmed.  The
problem of dealing with new types of strange solutions will likely
grow for larger values of $s$ and $N$.  However, one can hope that a
pattern for the types of strange solutions (for given values of $s$
and $N$) will eventually emerge.

We have exhibited some strange solutions in Tables
\ref{table:strangespin1} and \ref{table:strangespi32}.  Especially
notable is the solution $\{0\,, 0\}$ for $s=3/2\,, N=3$, which is
probably the simplest example of a solution with equal roots that
corresponds to an eigenstate of the Hamiltonian.

Several patterns can be discerned from Tables \ref{table:spin1},
\ref{table:spin32a}, \ref{table:spin32b}, similarly to the $s=1/2$
case \cite{Hao:2013jqa}.  In particular, for $M \sim s N$ and most
values of $N$, the number of unphysical singular solutions ${\cal
N}_{s}-{\cal N}_{sp}$ is comparable to $n(N,s N-M)$.  This suggests
that the naive formula (\ref{naive}) for the number of solutions of
the Bethe equations is incorrect not only for small values of $N$, but
also for $N \rightarrow \infty$.  This calls into question the proofs
of (\ref{naive}).  The related question \cite{Hao:2013jqa} of how singular
solutions fit within the string hypothesis also remains to be
clarified.  It is tempting to speculate that the number of strange
solutions in the spin-0 sector ($M = s N$) grows sufficiently fast in
the thermodynamic limit to account for the still-mysterious RSOS structure
\cite{Reshetikhin:1990jn} of the antiferromagnetic ground state.

\section*{Acknowledgments}
WH's research has been supported by the Mathematical Biosciences
Institute and the National Science Foundation under Grant DMS 0931642.
The work of RN was supported in part by the National Science
Foundation under Grant PHY-1212337, and by a Cooper fellowship.  WH
and AS also thank DARPA/AFRL Grant G-2457-2 for supporting
their work.

\appendix

\section{Singular solutions}\label{sec:singular}

We recall that the spin-$s$ Bethe equations (\ref{BAE}) admit solutions containing
\be
\{ i s\,, i(s-1)\,, \ldots \,, -i(s-1)\,, -i s \} \,,
\label{string}
\ee
which is a so-called exact string of length $2s+1$ centered at the
origin.  Since both the corresponding energy (\ref{energy}) and (with
our normalization of the $R$-matrix, see (\ref{Rmatrix}) below) the
corresponding Bethe state (\ref{vec}) are singular, we call such
solutions ``singular''.  It is always possible to regularize such
solutions in a way that the corresponding energies and Bethe states
are finite in the limit that the regulator is removed.  However, the
states obtained in this way are not guaranteed to be eigenstates of
the Hamiltonian.  We term ``physical'' a singular solution that leads
in this way to an eigenstate of the Hamiltonian; otherwise, we term
the singular solution ``unphysical''.  Generally, only a small
fraction of the singular solutions are physical.

We derive in this section a simple criterion for identifying the physical singular
solutions.  We proceed by generalizing the $s=1/2$ analysis in
\cite{Nepomechie:2013mua} to general values of $s$.

We use the standard $(\frac{1}{2},s)$ solution of 
the Yang-Baxter equation (see e.g. \cite{Babujian:1983ae})
\be
R^{(\frac{1}{2},s)}(\lambda) = \frac{1}{\lambda+i s} \left(\lambda + i \vec\sigma 
\dot \otimes \vec s \, \right) \,,
\label{Rmatrix}
\ee
except with an unusual normalization, such that the $R$-matrix is 
singular at $\lambda=-i s$. As in the Hamiltonian 
(\ref{Hamiltonian}), $\vec s$ are the spin-$s$ generators of 
$su(2)$; and $\vec \sigma$ are the standard Pauli matrices.
The monodromy matrix is given by
\be
T_{a}(\lambda) = R_{a N}^{(\frac{1}{2},s)}(\lambda) \cdots  
R_{a 1}^{(\frac{1}{2},s)}(\lambda) = \left(
\begin{array}{cc}
    A(\lambda) & B(\lambda) \\
    C(\lambda) & D(\lambda) 
\end{array}    \right) \,,
\ee 
and the transfer matrix is given by
\be
t(\lambda) = \tr_{a} T_{a}(\lambda) = A(\lambda) + D(\lambda) \,.
\ee
Its action on an off-shell 
Bethe vector (\ref{vec}) is given by  \cite{Faddeev:1996iy} 
\be
t(\lambda) |\lambda_{1} \,, \ldots \,, \lambda_{M} \rangle =
\Lambda(\lambda,\{\lambda\}) |\lambda_{1} \,, \ldots \,, \lambda_{M} \rangle +
\sum_{k=1}^{M} F_{k}(\lambda,\{\lambda\})\, 
|v_{k}(\lambda,\{\lambda\})\rangle  \,,
\label{offshell}
\ee
where
\be
|v_{k}(\lambda,\{\lambda\})\rangle \equiv B(\lambda)
\prod_{\scriptstyle{j \ne k}\atop \scriptstyle{j=1}}^M B(\lambda_{j}) 
|0\rangle \,,
\label{vk}
\ee
and
\be
\Lambda(\lambda,\{\lambda\}) &=& 
\prod_{j=1}^{M}\frac{\lambda-\lambda_{j}-i}{\lambda-\lambda_{j}}
+\left(\frac{\lambda-i s}{\lambda+i s}\right)^{N}
\prod_{j=1}^{M}\frac{\lambda-\lambda_{j}+i}{\lambda-\lambda_{j}}\,, 
\label{eigval}\\
F_{k}(\lambda,\{\lambda\}) &=& \frac{i}{\lambda-\lambda_{k}}\left[
\prod_{j\ne k}^{M}\frac{\lambda_{k}-\lambda_{j}-i}{\lambda_{k}-\lambda_{j}}
-\left(\frac{\lambda_{k}-i s}{\lambda_{k}+is} \right)^{N}
\prod_{j\ne k}^{M}\frac{\lambda_{k}-\lambda_{j}+i}{\lambda_{k}-\lambda_{j}} \right]
\,. \label{Fs}
\ee

If the set $\{ \lambda_{1} \,, \ldots \,, \lambda_{M} \} $ is not singular
(i.e., it does not contain the subset (\ref{string})), then the conditions
\be
F_{k}(\lambda,\{\lambda\}) = 0 \,, \qquad k = 1 \,, \ldots \,, M \,,
\label{BAE2}
\ee
imply that the ``unwanted'' terms in (\ref{offshell}) drop out; and 
hence the vector
$|\lambda_{1} \,, \ldots \,, \lambda_{M} \rangle $ is an eigenvector of the transfer 
matrix, with eigenvalue (\ref{eigval}).  Eqs.  (\ref{BAE2}) are
equivalent to the Bethe equations (\ref{BAE}).

We henceforth consider a solution $\{ \lambda_{1} \,, \ldots \,,
\lambda_{M} \} $ (where $M \ge 2s+1$) of the Bethe equations
(\ref{BAE}) that is singular, i.e. it {\em does} contain the subset (\ref{string}).  For
definiteness, we order the roots such that the first $2s+1$ roots are
given by (\ref{string}), respectively; that is,
\be
\lambda_{j} = i(s +1-j) \,, \qquad j = 1\,, \ldots\,, 2s+1 \,.
\label{gensing}
\ee 
We furthermore assume that there are no repeated roots. (Cases with 
repeated roots are considered in Section \ref{sec:strangesing}.)
That is, the remaining roots
$\lambda_{2s+2}\,, \ldots \,, \lambda_{M}$ are distinct and
are not equal to $is \,, i(s-1)\,, \ldots\,, - i(s-1)\,, -is $. We 
regularize this solution as follows
\be
\lambda_{j} = i(s +1-j) + \epsilon + c_{j} \epsilon^{N}\,, 
\qquad j = 1\,, \ldots\,, 2s+1 \,,
\label{gensingreg}
\ee 
where $\epsilon$ is the regulator, and $c_{1}\,, \ldots \,, c_{2s+1}$
are constants (independent of $\epsilon$) that are still to be
determined.  The key observation is that, due to our normalization 
of the $R$-matrix (\ref{Rmatrix}), $B(\lambda_{2s+1})$ is singular
\be
B(\lambda_{2s+1}) \sim \frac{1}{\epsilon^{N}} \,; \qquad 
B(\lambda_{k}) \sim 1  \,, \quad k = 1, \ldots, 2s \,,
\ee
as $\epsilon \rightarrow 0$.
Hence, in view of (\ref{offshell}), the vector $|\lambda_{1} \,, 
\ldots \,, \lambda_{M} \rangle $ is an eigenvector of the transfer 
matrix if \footnote{The vector $\lim_{\epsilon \rightarrow 0}\prod_{k=1}^{2s+1}B(\lambda_{k}) 
|0\rangle$ is finite, as shown for the case $s=1/2$ in \cite{Nepomechie:2013mua}.}
\be
F_{k}(\lambda,\{\lambda\}) \sim \epsilon^{N+1}  \,, \quad k = 1, \ldots, 2s \,; \qquad  
F_{2s+1}(\lambda,\{\lambda\}) \sim \epsilon 
\label{BAEgeneralized1}
\ee 
as $\epsilon \rightarrow 0$, together with 
\be
F_{k}(\lambda,\{\lambda\}) = 0 \,, \qquad k = 2s+2 \,, \ldots \,, M 
\,.
\label{BAEgeneralized2}
\ee 
We call (\ref{BAEgeneralized1}) generalized Bethe equations, since 
they are evidently generalizations of (\ref{BAE2}).

Evaluating $F_{k}(\lambda,\{\lambda\})$ using (\ref{Fs}) and 
(\ref{gensingreg}), the generalized Bethe equations 
(\ref{BAEgeneralized1}) imply
\be
c_{1}-c_{2} &=& 
\frac{(2s+1)}{(2is)^{N-1}}\prod_{j=2s+2}^{M}\frac{\lambda_{j}-i(s+1)}{\lambda_{j}-i(s-1)} \,, \non \\
c_{k}-c_{k+1} &=&  
-(c_{k-1}-c_{k})\left(\frac{1-k}{2s+1-k}\right)^{N-1}\left(\frac{2s+2-k}{k}\right)
\prod_{j=2s+2}^{M}\frac{\lambda_{j}-i(s+2-k)}{\lambda_{j}-i(s-k)} \,, 
\non \\
& & \qquad\qquad\qquad\qquad\qquad\qquad\qquad\qquad\qquad\qquad\qquad\qquad k =2 \,, \ldots\,, 2s \,, \non \\
c_{2s}-c_{2s+1} &=& 
-\frac{(2s+1)}{(-2is)^{N-1}}\prod_{j=2s+2}^{M}\frac{\lambda_{j}+i(s+1)}{\lambda_{j}+i(s-1)} \,.
\label{ceqs}
\ee
Forming the product of the first $2s$ equations in (\ref{ceqs}), we obtain an 
expression for $c_{2s}-c_{2s+1}$, which is consistent with the final 
equation in (\ref{ceqs}) provided that
\be
\prod_{j=2s+2}^{M}\left(\frac{\lambda_{j}-is}{\lambda_{j}+is}\right)
\left(\frac{\lambda_{j}-i(s+1)}{\lambda_{j}+i(s+1)}\right) = (-1)^{2 s N} \,.
\label{result1}
\ee

The remaining roots $\lambda_{2s+2}\,, \ldots \,, \lambda_{M}$ do not need 
regularization. They obey the usual Bethe equations 
(\ref{BAEgeneralized2}), which imply
\be
\left( \frac{\lambda_{k} + is}{\lambda_{k} - is}\right)^{N-1} 
\left(\frac{\lambda_{k} - i(s+1)}{\lambda_{k} + i(s+1)} \right) = 
\prod_{\scriptstyle{j \ne k}\atop \scriptstyle{j=2s+2}}^M 
\frac{\lambda_{k} -\lambda_{j} +i}{\lambda_{k} -\lambda_{j} -i}
\,, \qquad k = 2s+2 \,, \ldots \,, M \,.
\ee 
Forming the product of these equations, we obtain
\be
\prod_{k=2s+2}^{M}\left( \frac{\lambda_{k} + is}{\lambda_{k} - is}\right)^{N-1} 
\left(\frac{\lambda_{k} - i(s+1)}{\lambda_{k} + i(s+1)} \right)= 1 \,.
\label{result2}
\ee 

Combining the results (\ref{result1}) and (\ref{result2}), we arrive 
at the constraint
\be
\left[(-1)^{2s} \prod_{k=2s+2}^{M}\left( \frac{\lambda_{k} + 
is}{\lambda_{k} - is}\right) \right]^{N} = 1 \,.
\label{physical}
\ee
This is the condition for the singular solution (\ref{gensing}) with
all roots distinct to be physical.  For the case $s=1/2$, this
equation reduces to the one found in \cite{Nepomechie:2013mua}.

\section{Strange solutions}\label{sec:strange}

The Bethe equations (\ref{BAE}) generally admit many solutions with repeated
roots.  For $s=1/2$ and (at least) $N \le 14$ \cite{Hao:2013jqa}
such solutions do not lead to eigenstates of the Hamiltonian,
i.e. such solutions are unphysical.  However, for $s>1/2$, there exist
``strange'' solutions of the Bethe equations with repeated roots
that do produce eigenstates of the Hamiltonian, i.e. such solutions
are physical.  Here we discuss simple criteria for
determining if a given solution with repeated roots is physical.
We consider separately the cases that the solutions are 
singular.

\subsection{Strange non-singular solutions}\label{sec:strangenonsing}

We first consider solutions that are not singular (i.e., do not 
contain (\ref{string})) and contain a repeated root. In the simplest 
such case, a root $\mu$ has multiplicity 2. That is, the solution has the form
\be
\{\mu\,, \mu\,, \lambda_{3}\,, \ldots 
\,, \lambda_{M} \} \,,
\ee 
where $\mu\,, \lambda_{3}\,, \ldots \,, \lambda_{M}$ are distinct and are 
not equal to $is \,, i(s-1)\,, \ldots\,, - i(s-1)\,, -is $. Following 
\cite{Avdeev:1985cx, Izergin:1982hy}, one finds that
\be
\lefteqn{t(\lambda) |\mu \,, \mu\,, \lambda_{3}\,, \ldots \,, \lambda_{M} 
\rangle =
\Lambda(\lambda,\mu,\{\lambda\}) |\mu \,, \mu\,, \lambda_{3}\,, \ldots \,, 
\lambda_{M} \rangle \non} \\
&&
+\sum_{k=3}^{M} F_{k}(\lambda,\mu,\{\lambda\})\, 
B(\lambda)\, B(\mu)^{2}\prod_{\scriptstyle{j \ne k}\atop \scriptstyle{j=3}}^M B(\lambda_{j}) 
|0\rangle  
+
G(\lambda,\mu,\{\lambda\})\, 
B(\lambda)\, B'(\mu)\prod_{j=3}^M B(\lambda_{j}) |0\rangle
\non \\
&&+
H(\lambda,\mu,\{\lambda\})\, 
B(\lambda)\, B(\mu)\prod_{j=3}^M B(\lambda_{j}) |0\rangle 
\,, \label{ABArepeated}
\ee
where $\Lambda(\lambda,\mu,\{\lambda\})$ and $F_{k}(\lambda,\mu,\{\lambda\})$ 
are given by (\ref{eigval}) and (\ref{Fs}) with 
$\lambda_{1}=\lambda_{2}=\mu$, respectively; and 
\be
G(\lambda,\mu,\{\lambda\}) &=&
-\frac{1}{\lambda-\mu}\left[
\prod_{j=3}^{M}\frac{\mu-\lambda_{j}-i}{\mu-\lambda_{j}}
+\left(\frac{\mu-i s}{\mu+is} \right)^{N}
\prod_{j=3}^{M}\frac{\mu-\lambda_{j}+i}{\mu-\lambda_{j}} \right] \,.
\label{Gdef} \\
H(\lambda,\mu,\{\lambda\}) &=&
\frac{i}{\lambda-\mu}\Bigg\{\left(2-\frac{i}{\lambda-\mu}\right)
\prod_{j=3}^{M}\frac{\mu-\lambda_{j}-i}{\mu-\lambda_{j}}
-\left(2+\frac{i}{\lambda-\mu}\right)
\left(\frac{\mu-i s}{\mu+is} \right)^{N}
\prod_{j=3}^{M}\frac{\mu-\lambda_{j}+i}{\mu-\lambda_{j}}
\non\\
& &-i\frac{\partial}{\partial\mu}\left[
\prod_{j=3}^{M}\frac{\mu-\lambda_{j}-i}{\mu-\lambda_{j}}
+\left(\frac{\mu-i s}{\mu+is} \right)^{N}
\prod_{j=3}^{M}\frac{\mu-\lambda_{j}+i}{\mu-\lambda_{j}}
\right]\Bigg\} \,.
\label{Hdef}
\ee
Hence, besides the ``usual'' Bethe equations
\be
G(\lambda,\mu,\{\lambda\})  &=& 0 \,, \label{Geqn} \\
F_{k}(\lambda,\mu,\{\lambda\}) &=& 0\,, \qquad k = 3\,, \ldots \,, M\,,
\ee
the additional constraint
\be
H(\lambda,\mu,\{\lambda\})  = 0 \
\label{Heqn}
\ee
is necessary for all the unwanted terms in (\ref{ABArepeated}) to drop out.
Eq. (\ref{Geqn}) implies that the $\lambda$-dependent terms within braces in 
$H(\lambda,\mu,\{\lambda\})$ (\ref{Hdef}) cancel. Eq. (\ref{Heqn}) 
can therefore be rewritten as
\be
\hspace{-0.2in}& {\displaystyle \frac{\partial}{\partial \lambda}\left[
\left(\lambda-\mu-i\right)^{2}
\prod_{j=3}^{M}\frac{\lambda-\lambda_{j}-i}{\lambda-\lambda_{j}}
+\left(\lambda-\mu+i\right)^{2}
\left(\frac{\lambda-i s}{\lambda+i s}\right)^{N} 
\prod_{j=3}^{M}\frac{\lambda-\lambda_{j}+i}{\lambda-\lambda_{j}}
\right]\Bigg\vert_{\lambda=\mu}=0 \,.} 
\label{H2}
\ee

More generally, a root $\mu$ can have multiplicity $k$.  That is, the
solution has the form
\be
\{\underbrace{\mu\,, \ldots\,, \mu}_{k}\,, \lambda_{k+1}\,, \ldots 
\,, \lambda_{M} \} \,,
\label{strangenonsingular}
\ee 
where $\mu\,, \lambda_{k+1}\,, \ldots \,, \lambda_{M}$ are distinct and are 
not equal to $is \,, i(s-1)\,, \ldots\,, - i(s-1)\,, -is $. In order 
to derive all the physical constraints, one 
can proceed as above using the algebraic Bethe ansatz. A simpler
way is based on the observation that the Bethe
equations (\ref{BAE}) are the conditions for the transfer matrix
eigenvalue (\ref{eigval}) to be finite at $\lambda=\lambda_{k}$.
Similarly, for a solution of the form (\ref{strangenonsingular}), the transfer
matrix eigenvalue (\ref{eigval}) becomes
\be
\Lambda = 
\left(\frac{\lambda-\mu-i}{\lambda-\mu}\right)^{k}
\prod_{j=k+1}^{M}\frac{\lambda-\lambda_{j}-i}{\lambda-\lambda_{j}}
+\left(\frac{\lambda-\mu+i}{\lambda-\mu}\right)^{k}
\left(\frac{\lambda-i s}{\lambda+i s}\right)^{N} 
\prod_{j=k+1}^{M}\frac{\lambda-\lambda_{j}+i}{\lambda-\lambda_{j}}\,. 
\ee
Demanding that $\Lambda$ be finite at $\lambda=\mu$ leads to 
the following constraints
\be
& {\displaystyle \frac{\partial^{l}}{\partial \lambda^{l}}\left[
\left(\lambda-\mu-i\right)^{k}
\prod_{j=k+1}^{M}\frac{\lambda-\lambda_{j}-i}{\lambda-\lambda_{j}}
+\left(\lambda-\mu+i\right)^{k}
\left(\frac{\lambda-i s}{\lambda+i s}\right)^{N} 
\prod_{j=k+1}^{M}\frac{\lambda-\lambda_{j}+i}{\lambda-\lambda_{j}}
\right]\Bigg\vert_{\lambda=\mu}=0 \,,} \non \\
& l = 0\,, 1\,, \ldots \,, \,, k-1 \,. \label{extrastrangenonsingular}
\ee
Together with the usual Bethe equations for $\lambda_{k+1}\,, \ldots 
\,, \lambda_{M}$, these are the conditions for the solution 
(\ref{strangenonsingular}) to be 
physical. For the case $k=2$, Eqs. (\ref{extrastrangenonsingular}) 
with $l=0\,, 1$ are equivalent to Eqs. (\ref{Geqn}), (\ref{H2}), 
respectively.

\subsection{Strange singular solutions}\label{sec:strangesing}

We now consider solutions that are singular (i.e., contain 
(\ref{string})) and contain a repeated root. In principle, there 
are many such possibilities, which should be analyzed individually.
For simplicity, we explicitly analyze here only two such cases.

\subsubsection{$s=1$}\label{subsec:strange1}

We first consider the case that $s=1$
and the root 0 has multiplicity 2.  We order the roots such that the
first three are given by (\ref{string}), and the fourth root is the
second 0; i.e.
\be
(\lambda_{1}\,, \lambda_{2}\,, \lambda_{3}\,, \lambda_{4}) =
(i\,, 0\,, -i\,, 0 ) \,. 
\label{strange}
\ee
We assume that the remaining roots $\lambda_{5}\,, \ldots \,,
\lambda_{M}$ are distinct and are not equal to $\pm i \,, 0 $; they 
satisfy the usual Bethe equations (\ref{BAE}), which reduce in this 
case to
\be
\left( \frac{\lambda_{k} + i}{\lambda_{k} - i}\right)^{N-2} 
\left(\frac{\lambda_{k} - 2i}{\lambda_{k} + 2i} \right) = 
\prod_{\scriptstyle{j \ne k}\atop \scriptstyle{j=5}}^M 
\frac{\lambda_{k} -\lambda_{j} +i}{\lambda_{k} -\lambda_{j} -i}
\,, \qquad k = 5 \,, \ldots \,, M \,.
\ee 
We regulate this solution as follows
\be
\lambda_{1} &=& i  + \epsilon + c_{1} \epsilon^{N}\,, \non \\
\lambda_{2} &=& \epsilon + c_{2} \epsilon^{\frac{N}{2}}\,, \non \\
\lambda_{3} &=& -i  + \epsilon + c_{3} \epsilon^{N} \,, \non \\
\lambda_{4} &=& \epsilon + c_{4} \epsilon^{\frac{N}{2}}\,,
\label{strangereg}
\ee
where $\epsilon$ is the regulator, and $c_{1}\,, \ldots \,, c_{4}$
are constants (independent of $\epsilon$) that are still to be
determined. Note that the repeated roots are regulated with
$\epsilon^{\frac{N}{2}}$ instead of $\epsilon^{N}$. Note also that 
the repeated roots are no longer equal once they are regulated; 
hence, (\ref{ABArepeated}) should not be applied.

The generalized Bethe equations (\ref{BAEgeneralized1}) remain
valid for the roots that are not repeated, namely $k=1, 3$. They imply
\be
c_{2} c_{4} &= 
&-\frac{12}{(2i)^{N}}\prod_{j=5}^{M}\left(\frac{\lambda_{j}-2i}{\lambda_{j}}\right) \,, \non \\
c_{2} c_{4} &= 
&-\frac{12}{(-2i)^{N}}\prod_{j=5}^{M}\left(\frac{\lambda_{j}+2i}{\lambda_{j}}\right) \,,
\label{c2c4}
\ee
respectively, whose consistency requires
\be
\prod_{j=5}^{M}\left(\frac{\lambda_{j}+2i}{\lambda_{j}-2i}\right) = 
(-1)^{N} \,.
\label{strangespin1}
\ee 
This is the condition for the singular solution (\ref{strange}) to be
physical.\footnote{Solutions of the type (\ref{strange}) were already considered in
\cite{Avdeev:1985cx} (see Eq.  (18)).  However, a different physicality
condition was proposed there (see Eq.  (19)).  We believe that the latter is
incorrect, since it is based on (\ref{ABArepeated}), (\ref{H2}), which
should not be applied to singular solutions.\label{footnote:AV}}

We note that the generalized Bethe equation (\ref{BAEgeneralized1}) 
for the repeated root ($k=2$) is no longer valid. Indeed, for 
$\epsilon \rightarrow 0$, we observe that \footnote{We have checked 
(\ref{v2v4}) explicitly for $N=4$ and $N=6$.}
\be
|v_{2}(\lambda,\{\lambda\})\rangle &=& c_{4} |w\rangle + O(1) \,,  \non \\
|v_{4}(\lambda,\{\lambda\})\rangle &=& c_{2} |w\rangle + O(1) \,,  
\label{v2v4}
\ee
where $|v_{k}(\lambda,\{\lambda\})\rangle$ is defined in (\ref{vk}), 
and $|w\rangle \sim \epsilon^{-N/2}$.
Therefore, the unwanted terms in (\ref{offshell}) for $k=2$ and $k=4$ 
are not linearly independent,
\be
F_{2}(\lambda,\{\lambda\}) |v_{2}(\lambda,\{\lambda\})\rangle + F_{4}(\lambda,\{\lambda\}) 
|v_{4}(\lambda,\{\lambda\})\rangle = \left[ c_{4} F_{2}(\lambda,\{\lambda\}) + c_{2} 
F_{4}(\lambda,\{\lambda\}) \right] |w\rangle + O(1) \,. \non \\
\ee 
This implies the generalized Bethe equation
\be
c_{4} F_{2}(\lambda,\{\lambda\}) + c_{2} 
F_{4}(\lambda,\{\lambda\}) \sim \epsilon^{\frac{N}{2}+1} \,,
\ee  
which relates $c_{1}-c_{3}$ to $c_{2} c_{4}$.
Imposing the vanishing of the $O(1)$ terms leads to a further 
constraint.

\subsubsection{$s=3/2$}\label{subsec:strange2}

We now consider the case that $s=3/2$
and both roots $i/2$ and $-i/2$ have multiplicity 2.  \footnote{This
possibility was considered but rejected (prematurely, in our opinion)
in \cite{Avdeev:1985cx} below Eq.  (20).  As already noted, we believe that
(\ref{extrastrangenonsingular}) should not be applied to singular solutions.}
We order the roots such that 
\be
(\lambda_{1}\,, \ldots \lambda_{6}) =
(\frac{3i}{2}\,, \frac{i}{2}\,, -\frac{i}{2}\,, -\frac{3i}{2}\,, \frac{i}{2}\,, -\frac{i}{2}\ ) \,. 
\label{strange2}
\ee
We assume that the remaining roots $\lambda_{7}\,, \ldots \,,
\lambda_{M}$ are distinct and are not equal to $\pm 3i/2 \,, \pm i/2 $; they 
satisfy the usual Bethe equations (\ref{BAE}), which reduce in this 
case to
\be
\left( \frac{\lambda_{k} + \frac{3i}{2}}{\lambda_{k} - \frac{3i}{2}}\right)^{N-2} 
\left(\frac{\lambda_{k} - \frac{5i}{2}}{\lambda_{k} + \frac{5i}{2}} \right)
\left(\frac{\lambda_{k} - \frac{i}{2}}{\lambda_{k} + \frac{i}{2}} \right)= 
\prod_{\scriptstyle{j \ne k}\atop \scriptstyle{j=7}}^M 
\frac{\lambda_{k} -\lambda_{j} +i}{\lambda_{k} -\lambda_{j} -i}
\,, \qquad k = 7 \,, \ldots \,, M \,.
\ee 
Proceeding as in Sec. \ref{subsec:strange1}, we regulate this solution as follows
\be
\lambda_{1} &=& \frac{3i}{2} + \epsilon + c_{1} \epsilon^{N}\,, \non \\
\lambda_{2} &=& \frac{i}{2} + \epsilon + c_{2} \epsilon^{\frac{N}{2}}\,, \non \\
\lambda_{3} &=& -\frac{i}{2}  + \epsilon + c_{3} \epsilon^{\frac{N}{2}} \,, \non \\
\lambda_{4} &=& -\frac{3i}{2} +\epsilon + c_{4} \epsilon^{N}\,, \non \\
\lambda_{5} &=& \frac{i}{2} + \epsilon + c_{5} \epsilon^{\frac{N}{2}}\,, \non \\
\lambda_{6} &=& -\frac{i}{2}  + \epsilon + c_{6} \epsilon^{\frac{N}{2}} \,, 
\label{strangereg2}
\ee
where $\epsilon$ is the regulator, and $c_{1}\,, \ldots \,, c_{6}$
are constants (independent of $\epsilon$) that are still to be
determined. 

The generalized Bethe equations $F_{1} \sim \epsilon^{N+1}$ and  $F_{4} \sim \epsilon$
imply
\be
c_{2} c_{5} &= 
&\frac{8}{(3i)^{N-2}}\prod_{j=7}^{M}\left(\frac{\lambda_{j}-\frac{5i}{2}}{\lambda_{j}-\frac{i}{2}}\right) \,, \non \\
c_{3} c_{6} &= 
&\frac{8}{(-3i)^{N-2}}\prod_{j=7}^{M}\left(\frac{\lambda_{j}+\frac{5i}{2}}{\lambda_{j}+\frac{i}{2}}\right) \,,
\label{c2c5c3c6}
\ee
respectively. In contrast with (\ref{c2c4}), these relations do not 
immediately yield a consistency condition analogous to (\ref{strangespin1}).
Assuming relations similar to (\ref{v2v4}) for the unwanted terms 
associated with each set of repeated roots, we obtain
\be
F_{2}(\lambda,\{\lambda\}) |v_{2}(\lambda,\{\lambda\})\rangle + F_{5}(\lambda,\{\lambda\}) 
|v_{5}(\lambda,\{\lambda\})\rangle &=& \left[ c_{5} F_{2}(\lambda,\{\lambda\}) + c_{2} 
F_{5}(\lambda,\{\lambda\}) \right] |w\rangle + O(1) \,, \non \\
F_{3}(\lambda,\{\lambda\}) |v_{3}(\lambda,\{\lambda\})\rangle + F_{6}(\lambda,\{\lambda\}) 
|v_{6}(\lambda,\{\lambda\})\rangle &=& \left[ c_{6} 
F_{3}(\lambda,\{\lambda\}) + c_{3} 
F_{6}(\lambda,\{\lambda\}) \right] |w'\rangle + O(1) \,, \non \\
\label{assume}
\ee 
which imply the corresponding constraints 
\be
c_{5} F_{2}(\lambda,\{\lambda\}) + c_{2} 
F_{5}(\lambda,\{\lambda\})  &\sim & \epsilon^{\frac{N}{2}+1} \,, \non \\
c_{6} F_{3}(\lambda,\{\lambda\}) + c_{3} 
F_{6}(\lambda,\{\lambda\}) &\sim & \epsilon^{\frac{N}{2}+1} \,. 
\ee  
These constraints lead to expressions for $c_{1}$ and $c_{4}$ in 
terms of $c_{2} c_{5}$ and $c_{3} c_{6}$. 

In order to derive the sought-after consistency condition, we need a
relation between $c_{2} c_{5}$ and $c_{3} c_{6}$ in (\ref{c2c5c3c6}),
which presumably can be obtained by imposing the vanishing of the
$O(1)$ terms in (\ref{assume}).  Unfortunately, a direct computation
of these terms for the first nontrivial case ($N=8$) is out of reach.


\begin{thebibliography}{10}

\bibitem{Faddeev:1996iy}
L.~D. Faddeev, ``{How algebraic Bethe ansatz works for integrable models},'' in
  {\em Sym\'etries Quantiques (Les Houches Summer School Proceedings vol 64)},
  A.~Connes, K.~Gawedzki, and J.~Zinn-Justin, eds., pp.~149--219.
\newblock North Holland, 1998.
\newblock
\href{http://arxiv.org/abs/hep-th/9605187}{{\ttfamily arXiv:hep-th/9605187
  [hep-th]}}.
\newblock

\bibitem{Zamolodchikov:1980ku}
A.~Zamolodchikov and V.~Fateev, ``{Model factorized S matrix and an integrable
  Heisenberg chain with spin 1},''
{\em Sov.J.Nucl.Phys.} {\bfseries 32} (1980) 298--303.

\bibitem{Kulish:1981gi}
P.~Kulish, N.~Y. Reshetikhin, and E.~Sklyanin, ``{Yang-Baxter Equation and
  Representation Theory. 1.},''
\href{http://dx.doi.org/10.1007/BF02285311}{{\em Lett.Math.Phys.} {\bfseries 5}
  (1981) 393--403}.

\bibitem{Kulish:1981bi}
P.~Kulish and E.~Sklyanin, ``{Quantum spectral transform method. Recent
  developments},''
{\em Lect.Notes Phys.} {\bfseries 151} (1982) 61--119.

\bibitem{A.:1982zz}
L.~Takhtajan, ``{The picture of low-lying excitations in the isotropic
  Heisenberg chain of arbitrary spins},''
\href{http://dx.doi.org/10.1016/0375-9601(82)90764-2}{{\em Phys.Lett.}
  {\bfseries A87} (1982) 479--482}.

\bibitem{Babujian:1983ae}
H.~M. Babujian, ``{Exact solution of the isotropic Heisenberg chain with
  arbitrary spins: thermodynamics of the model},''
\href{http://dx.doi.org/10.1016/0550-3213(83)90668-5}{{\em Nucl.Phys.}
  {\bfseries B215} (1983) 317--336}.

\bibitem{Affleck:1988px}
I.~Affleck, D.~Gepner, H.~Schulz, and T.~Ziman, ``{Critical behavior of spin s
  Heisenberg antiferromagnetic chains: analytic and numerical results},''
\href{http://dx.doi.org/10.1088/0305-4470/22/5/015}{{\em J.Phys.} {\bfseries
  A22} (1989) 511}.

\bibitem{Reshetikhin:1990jn}
N.~Reshetikhin, ``{S matrices in integrable models of isotropical magnetic
  chains. 1.},''
{\em J.Phys.} {\bfseries A24} (1991) 3299--3310.

\bibitem{Bethe:1931hc}
H.~Bethe, ``{On the theory of metals. 1. Eigenvalues and eigenfunctions for the
  linear atomic chain},''
{\em Z.Phys.} {\bfseries 71} (1931) 205--226.

\bibitem{Takahashi1971}
M.~Takahashi, ``{One-dimensional Heisenberg model at finite temperature},''
  {\em Prog.Theor.Phys.} {\bfseries 46} (1971) 401--415.

\bibitem{Kirillov:1985}
A.~N. Kirillov, ``{Combinatorial identities, and completeness of eigenstates
  for the Heisenberg magnet},'' {\em J. Sov. Math.} {\bfseries 30} (1985) 2298.

\bibitem{Langlands:1995}
R.~P. Langlands and Y.~Saint-Aubin,
  \href{http://dx.doi.org/10.1007/3-540-59163-X_254}{``Algebro-geometric
  aspects of the {B}ethe equations,''} in {\em Strings and symmetries
  ({I}stanbul, 1994)}, vol.~447 of {\em Lecture Notes in Phys.}, pp.~40--53.
\newblock Springer, Berlin, 1995.

\bibitem{Langlands:1997}
R.~P. Langlands and Y.~Saint-Aubin, ``Aspects combinatoires des \'equations de
  {B}ethe,'' in {\em Advances in mathematical sciences: {CRM}'s 25 years
  ({M}ontreal, {PQ}, 1994)}, L.~Vinet, ed., vol.~11 of {\em CRM Proc. Lecture
  Notes}, pp.~231--301.
\newblock Amer. Math. Soc., Providence, RI, 1997.

\bibitem{Fabricius:2000yx}
K.~Fabricius and B.~M. McCoy, ``{Bethe's equation is incomplete for the XXZ
  model at roots of unity},''
  \href{http://dx.doi.org/10.1023/A:1010380116927}{{\em J.Statist.Phys.}
  {\bfseries 103} (2001) 647--678},
\href{http://arxiv.org/abs/cond-mat/0009279}{{\ttfamily arXiv:cond-mat/0009279
  [cond-mat.stat-mech]}}.

\bibitem{Baxter:2001sx}
R.~J. Baxter, ``{Completeness of the Bethe ansatz for the six and eight vertex
  models},'' \href{http://dx.doi.org/10.1023/A:1015437118218}{{\em
  J.Statist.Phys.} {\bfseries 108} (2002) 1--48},
\href{http://arxiv.org/abs/cond-mat/0111188}{{\ttfamily arXiv:cond-mat/0111188
  [cond-mat]}}.

\bibitem{Mukhin:2009}
E.~Mukhin, V.~Tarasov, and A.~Varchenko, ``{Bethe algebra of homogeneous XXX
  Heisenberg model has simple spectrum},'' {\em Commun. Math. Phys.} {\bfseries
  288} (2009) 1--42, \href{http://arxiv.org/abs/0706.0688}{{\ttfamily
  arXiv:0706.0688 [math]}}.

\bibitem{Avdeev:1985cx}
L.~V. Avdeev and A.~A. Vladimirov, ``{Exceptional solutions of the Bethe ansatz
  equations},''
\href{http://dx.doi.org/10.1007/BF01037864}{{\em Theor. Math. Phys.} {\bfseries
  69} (1987) 1071}.

\bibitem{Essler:1991wd}
F.~H.~L. Essler, V.~E. Korepin, and K.~Schoutens, ``{Fine structure of the
  Bethe ansatz for the spin 1/2 Heisenberg XXX model},''
{\em J.Phys.} {\bfseries A25} (1992) 4115--4126.

\bibitem{Siddharthan:1998}
R.~Siddharthan, ``{Singularities in the Bethe solution of the XXX and XXZ
  Heisenberg spin chains},''
  \href{http://arxiv.org/abs/cond-mat/9804210}{{\ttfamily
  arXiv:cond-mat/9804210 [cond-mat]}}.

\bibitem{Noh:2000}
J.~D. Noh, D.-S. Lee, and D.~Kim, ``{Origin of the singular Bethe ansatz
  solutions for the Heisenberg XXZ spin chain},'' {\em Physica A} {\bfseries
  287} (2000) 167.

\bibitem{Beisert:2003xu}
N.~Beisert, J.~A. Minahan, M.~Staudacher, and K.~Zarembo, ``{Stringing spins
  and spinning strings},'' {\em JHEP} {\bfseries 0309} (2003) 010,
\href{http://arxiv.org/abs/hep-th/0306139}{{\ttfamily arXiv:hep-th/0306139
  [hep-th]}}.

\bibitem{Beisert:2004hm}
N.~Beisert, V.~Dippel, and M.~Staudacher, ``{A Novel long range spin chain and
  planar N=4 super Yang-Mills},''
  \href{http://dx.doi.org/10.1088/1126-6708/2004/07/075}{{\em JHEP} {\bfseries
  0407} (2004) 075},
\href{http://arxiv.org/abs/hep-th/0405001}{{\ttfamily arXiv:hep-th/0405001
  [hep-th]}}.

\bibitem{Hagemans:2007}
R.~Hagemans and J.-S. Caux, ``{Deformed strings in the Heisenberg model},''
  {\em J.Phys.A} {\bfseries 40} (2007) 14605,
  \href{http://arxiv.org/abs/0707.2803}{{\ttfamily arXiv:0707.2803
  [cond-mat]}}.

\bibitem{Arutyunov:2012tx}
G.~Arutyunov, S.~Frolov, and A.~Sfondrini, ``{Exceptional Operators in N=4
  super Yang-Mills},'' \href{http://dx.doi.org/10.1007/JHEP09(2012)006}{{\em
  JHEP} {\bfseries 1209} (2012) 006},
\href{http://arxiv.org/abs/1205.6660}{{\ttfamily arXiv:1205.6660 [hep-th]}}.

\bibitem{Nepomechie:2013mua}
R.~I. Nepomechie and C.~Wang, ``{Algebraic Bethe ansatz for singular
  solutions},'' \href{http://dx.doi.org/10.1088/1751-8113/46/32/325002}{{\em
  J.Phys.} {\bfseries A46} (2013) 325002},
\href{http://arxiv.org/abs/1304.7978}{{\ttfamily arXiv:1304.7978 [hep-th]}}.

\bibitem{Hao:2013jqa}
W.~Hao, R.~I. Nepomechie, and A.~J. Sommese, ``{Completeness of solutions of
  Bethe's equations},'' {\em Phys.Rev.E} {\bfseries 88} (2013) 052113,
\href{http://arxiv.org/abs/1308.4645}{{\ttfamily arXiv:1308.4645 [math-ph]}}.

\bibitem{SW2}
A.~J. Sommese and C.~W. Wampler, {\em The Numerical Solution Of Systems Of
  Polynomials Arising In Engineering And Science}.
\newblock World Scientific Publishing Company Incorporated, 2005.

\bibitem{BHSW1}
D.~J. Bates, J.~D. Hauenstein, A.~J. Sommese, and C.~W. Wampler, {\em
  Numerically Solving Polynomial Systems with Bertini}.
\newblock SIAM, 2013.

\bibitem{Mendonca:2013}
J.~R.~G. Mendon\c{c}a, ``{Exact eigenspectrum of the symmetric simple exclusion
  process on the complete, complete bipartite, and related graphs},'' {\em
  J.Phys.} {\bfseries A46} (2013) 295001,
  \href{http://arxiv.org/abs/1207.4106}{{\ttfamily arXiv:1207.4106
  [cond-mat.stat-mech]}}.

\bibitem{Bollinger:1993}
R.~C. Bollinger, ``{Extended Pascal triangles},'' {\em Mathematics Magazine}
  {\bfseries 66} (1993) 87--94.

\bibitem{tables}
See associated supplemental material on the arxiv.

\bibitem{Izergin:1982hy}
A.~Izergin and V.~Korepin, ``{Pauli principle for one-dimensional bosons and
  the algebraic Bethe ansatz},''
\href{http://dx.doi.org/10.1007/BF00400323}{{\em Lett.Math.Phys.} {\bfseries 6}
  (1982) 283--288}.

\end{thebibliography}

\providecommand{\href}[2]{#2}\begingroup\raggedright\endgroup

\end{document}